# Self-Assembled Biosensors on a Solid Interface for Rapid Detection and Growth Monitoring of Bacteria

*Kinnunen, P\*, M. E. Carey, E. Craig, S. Brahmasandra, B. H. McNaughton\**

*Life Magnetics, Inc., 1600 Huron Parkway, Building 520, Ann Arbor, MI 48109, US*

*Email: pkkinn@umich.edu, mcarey@lifemagnetics.com, ecraig@lifemagnetics.com, sundu@lifemagnetics.com, brandon@lifemagnetics.com* \*Contact authors

**Abstract**

Developing rapid methods for pathogen detection and growth monitoring at low cell and analyte concentrations is an important goal, which numerous technologies are working towards solving. Rapid biosensors have already made a dramatic impact on improving patient outcomes and with continued development, these technologies may also help limit the emergence of antimicrobial resistance and reduce the ever expanding risk of foodborne illnesses. One technology that is being developed with these goals in mind is asynchronous magnetic bead rotation (AMBR) biosensors. Self-assembled AMBR biosensors have been demonstrated at water/air and water/oil interfaces, and here, for the first time, we report on self-assembled AMBR biosensors used at a solid interface. The solid interface configuration was used to measure the growth of *Escherichia coli* with two distinct phenomena at low cell concentrations: firstly, the AMBR rotational period decreased and secondly, the rotational period increased after several division times. Taking advantage of this low cell concentration behavior, a 20 % signal change from the growth of *E. coli* O157:H7 was detected in 91 ± 4 minutes, with a starting concentration of 5 x $10^3$ CFU/mL. Such a rapid cell growth sensor could dramatically improve the detection time and sensitivity in applications requiring phenotypic testing of target cells.

## 1. Introduction

Due to the recognized value in performing rapid and low concentration assays, many technologies have been emerging for characterizing cellular growth. Biosensor techniques for measuring bacterial growth at low concentrations include acoustic resonance (Chang et al., 2007), electrical impedance (Gómez et al., 2002), UV resonance Raman spectroscopy (Neugebauer et al., 2006), and differential calorimetry (Chang-Li et al., 1988). However, most of these techniques still suffer from relatively long turnaround times (4 hours or more) and are unable to perform at low starting cell concentrations ($10^4$ CFU/mL and less). As a result, there has been a continued need for the development of more sensitive ways to measure cellular growth, leading to the ultimate sensitivity limit of detecting and even monitoring the growth of individual bacterial cells (Burg et al., 2007; McNaughton et al., 2007; Godin et al., 2010; Kinnunen et al., 2011). Techniques that have been used to measure the growth of individual bacterial cells include cantilevers (Gfeller et al., 2005a, 2005b; Bryan et al., 2010; Godin et al., 2010), asynchronous magnetic bead rotation (Kinnunen et al., 2011), and optical microscopy (Boedicker et al., 2009). However, with these reported approaches, significant challenges remain in implementing the methods in a clinical setting. For example, cantilevers may require microfabrication and do not perform submerged in fluid, which can increase the complexity of creating and using such devices. Microscopy methods on the other hand can be labor intensive and potentially low throughput. AMBR biosensors have also been used to detect the growth of individual bacterial cells, but this was accomplished using a high cost and low throughput experimental setup (Kinnunen et al., 2011). As a result, large scale application of this single cell sensitivity could be challenging. Therefore, there remains a need for a rapid, robust, and scalable system for measuring bacterial growth at low numbers in a fluidic sample. The self-assembled AMBR biosensors on a solid interface—reported here—address this issue by using a highly sensitive approach in a standard disposable 384-microwell plate, making sample preparation and implementation straight-forward.

In recent years, there have been many advances in both genotypic and phenotypic bacterial detection technologies. Immunoassays have a long history of use for identifying bacterial species and have also been used for detecting bacterial resistance markers (e.g. the BinaxNow® from Inverness Medical Innovations, Inc., which is a PBP2a agglutination test to detect resistance in *Staphylococcus aureus*) (Romero-Gómez et al., 2012). Immunoassays have the advantage of being low-cost and easy to use (e.g. lateral flow assay); however, immunoassays typically suffer from drawbacks associated with low clinical sensitivity and specificity. Nucleic acid tests (NATs), such as real time PCR, have become pervasive in healthcare—especially for the identification of infectious agents or genetic fingerprints, such as antibiotic resistance profiles—and food testing applications due to their high sensitivity and specificity (Lazcka et al., 2007). However, NAT-based methods do not directly measure cellular growth. As a result, NATs currently lack the ability to identify the resistance of species where the resistance is expressed through multiple mechanisms—as is the case in Gram-negative bacteria—and the resistance where the fingerprint is not previously known or has changed due to the emergence of a new strain (Jorgensen and Ferraro, 1998, 2009). Several studies have explored the use of NAT to perform antimicrobial susceptibility testing (AST), obtaining the minimum inhibitory (MIC) concentrations of various antibiotics (Waldeisen et al., 2011). While these approaches have served as "proof of concept" demonstrations, they remain highly impractical for routine implementation due to the high cost and complexity of performing up to a 100 reactions to determine one MIC panel. This cost-related issue is one of the reasons why turbidity remains the gold standard for performing microbroth dilutions (Jorgensen and Ferraro, 2009). As a result, the motivation behind this study is to develop a technology that would combine the robustness and cost effectiveness of culture, the ease of use and simplicity of immunoassays, while approaching the speed of NAT techniques.

Previously published self-assembled AMBR biosensors have reported time-to-results for 1 x $10^5$ CFU/mL of *E. coli* O157 of approximately 4 hours of instrument time, where time-to-results is a 20% change in biosensor signal (Kinnunen et al., 2012) . Ideally, lower concentrations of bacteria could be detected with reduced time-to-results, allowing for a cell growth-based technique to provide results in comparable times to some reported NAT techniques. Most of the previously published AMBR methods utilize a hanging drop or water-in-oil drop to keep the beads from adhering to a solid interface (Sinn et al., 2011; Kinnunen et al., 2012). In order to make the AMBR method easier to implement and to improve the time-to-results, standard well plates were used in this study. As reported in this paper, when performing solid-interface AMBR, the rotational period decreases initially before increasing, see Fig. 1 for a schematic illustration. This initial decrease of the rotational period has not been observed at a solid interface for a self-assembled AMBR sensor before. Self-assembled AMBR biosensors at water/air interfaces do not exhibit this behavior, rather, the rotational period merely slows down due to bacterial growth and an increase in drag. The decrease of the rotational period at the solid interface happens earlier, and may therefore be utilized as a rapid bacterial growth indicator. Reductions in the rotational period upon bacterial growth, using single-bead AMBR biosensors, has been observed before; however, this was observed with high cell concentrations (Sinn et al., 2012). Furthermore, reductions in viscosity due to bacterial motility have also been reported at very high cell concentrations (eg. $10^{10}$ CFU/mL) (Sokolov and Aranson, 2009).

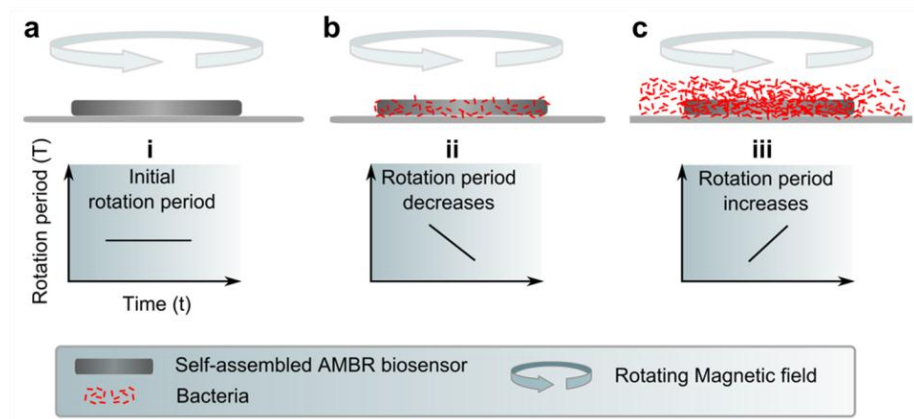

**Fig. 1**. An illustration of the rotational behavior of a self-assembled AMBR biosensor on a solid interface, with schematic figures on top and the corresponding rotational behavior on the bottom. (a) This is the initial state, where the AMBR group rotates with a constant rotation period on a solid interface. (b) As the bacteria start to grow, the rotational dynamics are changed and the rotational period of the AMBR group decreases. (c) After sufficient period of growth, the "traditional" increase can be seen in the AMBR rotational period, and this happens at a later point in time. Three separate phases i, ii, and iii have been identified, which correspond to parts a, b and c.

## 2. Materials and Methods

### 2.1 Bacterial culture

Bacterial isolates were obtained from ATCC and stored according to supplier recommendations. *E. coli* O157:H7 ATCC 35150 and *S. aureus* ATCC 29213 were used in the experiments. Before each experiment, 2-3 bacteria colonies were resuspended in CA-MHB, grown to log phase and diluted to 0.5 McFarland standard concentration, which roughly corresponds to $10^8$ CFU/mL. Bacteria were then diluted to the desired concentration. Cell counts were performed on MH agar plates to obtain approximate starting concentration values.

### 2.2 Sample preparation

Anti-*E. coli* O157 functionalized, 2.8 µm diameter magnetic particles (20 µl, Invitrogen 710-03) were mixed with bacteria diluted to 5 x $10^3$ CFU/ml (0.9 ml), and allowed to incubate (10 min at 37 °C) with end-over-end rotation. Washing of the particles was performed (Life Technologies, DynaMag-2) two times and the sample was resuspended in a final solution of CA-MHB (160 µl) that was spiked with 1% Pluronic F-127 surfactant to reduce adhesion between magnetic beads and the polystyrene solid-interface of the well plate. Samples were then pipetted into 384 microwell plate (18 µl each well, (Greiner Bio-One 788101)), and the plate was set on a permanent magnet array (VP scientific, Inc. #771BT-Q-1A) for 5 minutes to form magnetic groups (See section 2.3 for details). Plates were then placed in a custom made prototype device for observation (for images of the prototype see Supplementary data Fig. S2). In order to determine the starting concentration, the samples were diluted appropriately at the end of the sample preparation process, plated on a MH agar plate and the colonies were counted 18 hours later to obtain approximate bacterial concentrations and magnetic separation capture efficiencies. The total sample preparation time was roughly 40 minutes, which is not included in the graphs.

*2.3 Self-assembled magnetic bead group formation*

Stability of the magnetic bead group is one of the most critical steps of the whole process. Similar magnetic particle systems on solid interfaces in a rotating magnetic field have been studied (Nagaoka et al., 2005; Weddemann et al., 2010), where the formation of a single cohesive unit is usually something that is avoided. The cohesive groups, described here, were achieved by using round flat bottom polystyrene microplates, where each well was cylindrical with a flat bottom (Greiner Bio-One 788101). A schematic representation of the formation of the AMBR biosensors in a standard 384 well plate, using a permanent magnetic array (VP scientific, Inc. #771BT-Q-1A) can be seen in Fig. 2a. When forming the magnetic bead groups at the bottom of the well plate, the position of the magnet array is critical. Specifically, robust groups formed when the magnet was off-center from the well plate (see Figures 2a and Supplementary data Fig. S3), and to ensure reproducibility a micropositioner (Thorlabs MT-X) was used. Also, when removing the plate from the magnetic array, the plate was removed vertically so that groups did not reform during removal of the plate. Failing to follow this step might lead to false positive results. In an eventual system incorporating this approach, the magnets can be moved up and down automatically thereby making this a very reliable and robust process. Furthermore, this positioning helped form groups with optical asymmetry, which is needed for the prototype measurements. If using non-optical measurement methods like GMRs or Hall sensors, this asymmetry may not be required. Well plates holding the magnetic bead groups were then placed in a custom made prototype system (Fig. 2b).

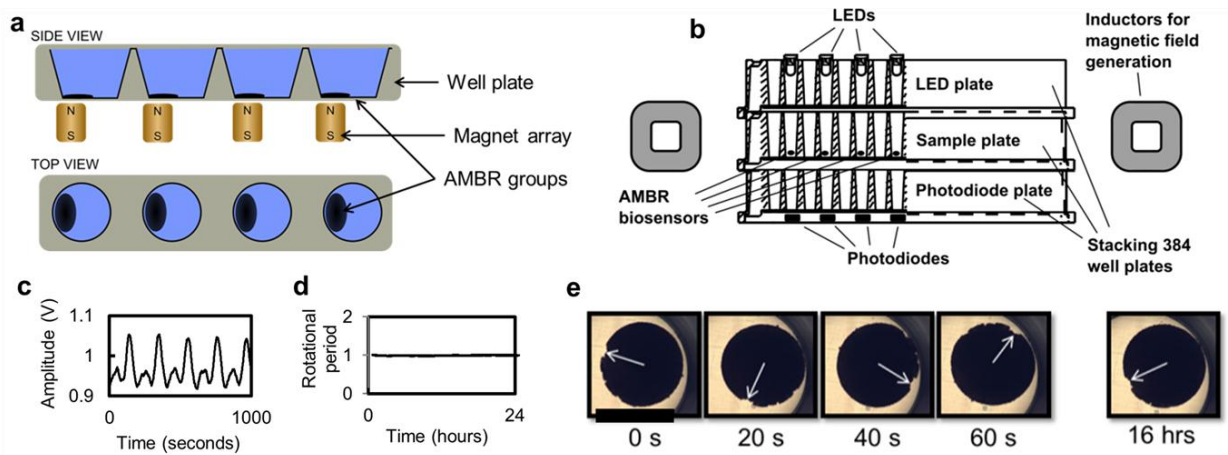

**Fig. 2.** Overview of the self-assembled AMBR biosensor formation and the prototype system for the biosensor monitoring. (a) Side and top view of the AMBR group formation process, where a magnet array was placed under the well plate for group formation. (b) Cross-sectional side view of the AMBR prototype, which consisted of three stackable 384-well plates. Every other well contains an AMBR group, the rotation of which was observed with an LED and a photodiode pair, connected to a computer. The top and bottom plate were used to house the LEDs and photodiodes. A rotating magnetic field was generated using an inductor setup around the plates. (c) Raw intensity data from a single prototype photodiode, where the rotation can be seen as a recurring peaks in the voltage. (d) The rotational period computed from the data in part c, using a tone measurement algorithm and normalized to one. (e) Microscope images of an AMBR group rotating in a single well in a 384 well plate with no bacteria. Note the constant shape of the group even after 16 hours. The scale bar is 1 mm.

*2.4 Prototype design and operation*

The prototype consisted of an LED (350-2318-ND) on top of each well and a photodiode (TSL257) underneath to measure fluctuation of the passed light due to the rotation of the AMBR biosensor. Cross section view of the prototype device can be seen in Fig. 2b, where the prototype consists of three stackable 384 microwell plates. The photodiodes are connected to a data acquisition board, which is connected to a computer and analyzed with custom software. Since AMBR biosensors can be actuated using a comparably low quality and low amplitude (<10 mT) rotating magnetic field with high tolerance in circularity and gradients, a low-cost and high-area magnetic field generation is possible. Around the stacked plates, shown in Fig. 2b, are the inductors (eight 10 mH I-core inductors, ERSE Audio, ELC54-19-10000) that are used for generating the rotating magnetic field. The top view of the inductor setup can be seen in Supplementary data (Fig. S4). By arranging eight inductors in the shown manner, and connecting Phase A to amplified sine-wave generator and Phase B to amplified cosine-wave (or vice versa), a large and uniform rotating magnetic field can be generated that covers the area of a standard microwell plate (7.9 ± 1 mT, 10 Hz). The inductors were actuated by a stepper motor driver (Gecko 201X), which was powered with a power supply (48V, 1A), and saw-tooth signal (5V, 2.5 mA) with 40 times the frequency needed for the rotating magnetic field, in this case 400 Hz saw tooth signal for 10 Hz rotating magnetic field. The prototype as constructed demonstrated high robustness—it was used to perform over a hundred experiments within seven months, and it was running over 5000 hours during that period.

*2.5 Data acquisition and analysis*

The signals from the photodiodes were acquired with four low cost data acquisition boards (NI 6008), using a custom written LabView (National Instruments, TX) program. The signals were acquired at 1 Hz, and the rotational period of the AMBR sensor in each well was calculated in real time using a subroutine for frequency analysis (Tone Measurements VI). An 1800 second data window was used for this analysis to ensure several rotational periods to occur within the data window. The initial AMBR rotational periods were usually between 100 and 300 seconds for these measurements; however it is highly dependent on the specifics of the group formation procedure and rotational periods as short as 10 seconds and as long as two hours were observed.

### 3. Results and discussion

Self-assembled AMBR biosensors were successfully implemented in a standard 384 well microplate, and observed using a custom made prototype. This was achieved using flat bottom polystyrene microplates, 2.8 μm magnetic microbeads and Mueller Hinton II broth for microbiology applications, spiked with a surfactant to reduce adhesion between magnetic beads and the polystyrene plate. In order for self-assembled AMBR measurements to be used, it is important for the magnetic group to stay together during rotation, for that rotation to be reproducible, and for the group to have enough optical asymmetry to create a modulated signal on the photodetector. All of these items were accomplished using the described methods and the prototype system. Also, the prototype was placed in a laboratory incubator (37 °C), and the AMBR biosensors were monitored with a connected PC. The raw voltage signal of the photodiode can be seen in Figure 2c. The signal was sufficient to use a standard tone algorithm to calculate the rotational period. This rotational period was tracked over a 24-hour period and the magnetic bead group rotation was highly reproducible, with a CV of 0.8%. The data shown in Figure 2c and 2d was generated by measuring the rotation of the magnetic bead group shown in Figure 2e. The shown magnetic bead group has approximately $10^6$ beads.

The self-assembled AMBR biosensor response to bacterial growth, when rotated on a solid interface, behaved differently than what is observed at air/water and oil/water interfaces. In previous self-assembled AMBR applications at air/water and oil/water interfaces the rotation period of the sensor increased in response to bacterial growth – however, as observed in this study, on a solid interface this was not the case. On a solid interface, the rotational period initially decreases in response to bacterial growth—to roughly half the initial rotational period—before starting to increase, see Fig. 3a. This effect was observed with bacterial concentrations between $5 \times 10^2$ and $5 \times 10^4$ CFU/mL. Figure 3a also shows the flat control data of a typical well that does not contain any bacteria. The control measurement was performed in the same plate as the growth measurements. With lethal concentrations of gentamicin (e.g. 2 μg/mL) in the growth media, there was no reduction in rotational period, further indicating that the reduction in the rotational period is due to bacterial growth, see Fig. 3b. Finally, the decrease of the rotational period occurred earlier in time than what has been previously observed with air/water and oil/water interface measurements, allowing for a significant reduction in instrument time from 240 min—using the "traditional" air/water interface AMBR (Kinnunen et al., 2012)—to 126 min using the described method. Additional reduction in time-to-results was achieved by using a larger initial volume leading to a time-to-results of $91 \pm 4$ min with 10 mL samples with *E. coli* O157:H7 at $5 \times 10^3$ CFU/mL, see Fig. 3c. A significant difference between bacterial growth and the lack of growth due to antibiotics was seen within 120 minutes (including sample preparation time, see Fig. 3b), which corresponds to $4 \times 10^4$ CFU/well ($1.6 \times 10^5$ CFU/mL) at the point of 20 % AMBR signal change. These results also suggest that the growth or lack of growth of a small concentration of cells can be rapidly detected using AMBR biosensors on a solid interface. More research is needed to fully characterize the mechanism of the decrease in rotational period which may relate to friction (Chang et al., 2009), drag (Agayan et al., 2008), or biofilm formation (Souza et al., 2010); however what we have determined from this investigation is that self-assembled AMBR can be performed on a solid interface with a motile gram negative bacterial model organism (*E. coli*) and a gram positive non-motile bacterial model organism *S. aureus* (Supplementary data, Fig. S5). We also established that the reduction in the rotational period was a result of bacterial growth. The goal of this research was to demonstrate self-assembled AMBR sensors on a solid interface and to report on the behavior of the growth of bacteria with this kind of sensor on a solid interface.

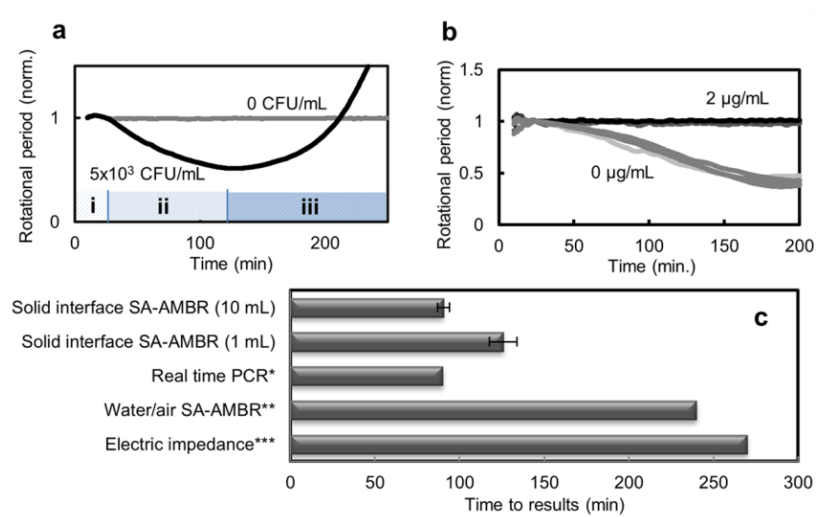

**Fig. 3.** The rotational period and instrument time for bacterial growth measurements in a standard 384 well plate, using an AMBR biosensor on a solid interface. (a) The rotational period of a self-assembled AMBR biosensor at a solid interface, inoculated with 5 x $10^3$ CFU/mL *E. coli*, 10 mL of sample. After a brief stationary phase (i) a reduction of the rotational period (ii) is observed, and after roughly two-fold reduction, the rotational period begins to increase (iii). The three distinct phases (i), (ii), and (iii) from Fig. 1 are shown above the x-axis. The data shown is an average of 6 wells, and the average instrument time for across all six wells was 51 ± 4 minutes with 40 minute sample preparation. The average initial rotational periods of 200 ± 28 seconds before normalization. Data before 10 minutes was omitted. (b) The rotational periods of self-assembled AMBR biosensors at a solid interface, with 5 x $10^3$ CFU/mL *E. coli*, 1 mL sample, with 0 and 2 μg/mL gentamicin antibiotic. Two μg/mL of gentamicin is above the MIC, and therefore no growth is observed. Four overlapping data sets of each condition are shown. (c) A time-to-results comparison for solid interface self-assembled AMBR with 1 mL and 10 mL sample, real time PCR, water/air self-assembled AMBR biosensor, and electric impedance measurement. *(Kinnunen et al., 2012), **(Mothershed and Whitney, 2006; "DuPont Qualicon BAX® detection system," 2012), ***(Gómez et al., 2002).

As mentioned, the decrease in the rotational period of the self-assembled AMBR sensors on solid interface was not restricted to *E. coli* O157:H7, or even to Gram-negative rods – similar decreases in the rotational period were observed with Gram-positive coccal *S. aureus* bacteria, which is shown in Supplementary data (Fig. S5). In order to determine that the early-time reduction in the rotational period was indeed caused by growth and not caused by a contraction of the magnetic bead group, a magnetic bead group was imaged using a low magnification optical microscope, housed with an incubator. The results confirm that while the rotational period decreased, the magnetic bead group did not contract in size (Supplementary data, Fig. S1).

The sensitivity of the self-assembled AMBR on solid interface technique may also enable novel applications. One such potential application is to use the method to perform antimicrobial susceptibility testing from urine samples or from positive blood cultures. While the use of immunomagnetic separation may not be practical for broad scale microbial identification, the AMBR biosensor described in this paper can be very easily implemented for specific target identification and determining its antimicrobial susceptibility. While additional studies need to be performed in order to fully characterize the phenomena observed here, we anticipate the possibility of potential implementation with other cell types such as mammalian, yeast, and fungi.

### 4. Conclusions

Self-assembled AMBR biosensors have previously been implemented at water/air interface, and here, for the first time, we report on the use of self-assembled AMBR biosensors at a solid interface. The solid interface configuration led to previously unobserved results, where the presence of low levels of bacteria led to rapid time-to-results via a reduction in the rotational period of the AMBR sensor. Taking advantage of this new observation, 5 x $10^3$ CFU/mL *E. coli* O157:H7 was detected in 91 ± 4 min with 10 ml of sample and in 126 ± 8 min with 1 ml of sample, including sample preparation and instrument time. The ability to rapidly measure cell growth at low concentrations may enable novel improvements to growth-based testing, such as antimicrobial susceptibility testing.

**Acknowledgements**

The authors would like to thank Betty Wu and Dr. Hao Xu for contributions in the lab that led to this project.

**Supplementary data**

See below for supplementary data and figures for "Self-Assembled AMBR Biosensors on a Solid Interface for Rapid Detection and Growth Monitoring of Bacteria".

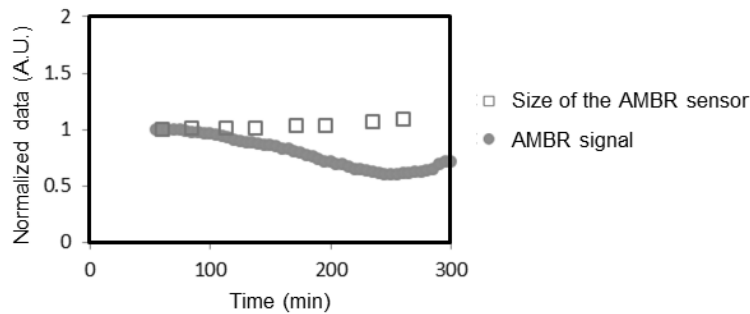

**Fig. S1.** To establish that the reduction in rotational period is not due to size changes, the size of a self-assembled AMBR biosensor on a solid interface was measured on a microscope, and compared to its rotational period. Normalized size of the biosensor is depicted in the plot with empty squares, and normalized rotational period with filled circles. The size of the biosensor does not decrease, when the rotational period decreases. This experiment was done at room temperature. The data is normalized at 50 minutes and the data prior to that is not shown.

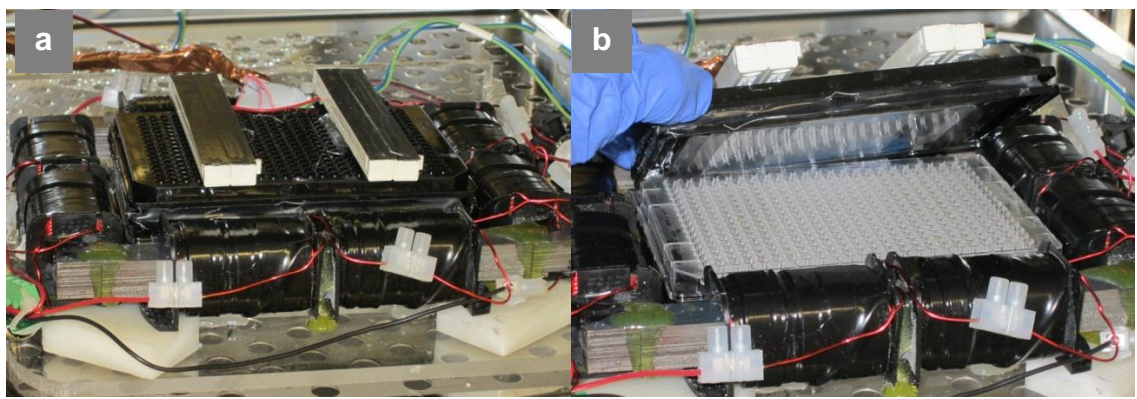

**Fig. S2.** Images of the prototype device in an incubator with (a) lid closed and (b) lid open showing the sample plate.

*More on self-assembled magnetic bead group formation*

The use of self-assembled AMBR biosensors was also enabled by using a surfactant in the surrounding medium at a sufficient concentration (i.e. Pluronic F127 at 1%), and by the use of cylindrical wells with a flat bottom. The surfactant aided in mitigating the adherence of the magnetic beads to the bottom of the well plate. This study did not investigate alternative surfaces, concentrations, or surfactants. Non-ionic surfactants such as Pluronic have been used as growth-promoting additives to animal and microbial cultures, especially in large batch reactors and can therefore be used without adversely affecting bacterial growth (Ntwampe et al., 2010). Also by subjecting the samples to a strong magnetic gradient (achieved by using an array of permanent magnets), magnetic beads assembled into cohesive groups in the MH broth, and possessed enough optical asymmetry for straight-forward measurement of the rotation on the prototype system or on an optical microscope.

Using AMBR on a solid interface can pose additional challenges when rotating a group of magnetic beads. For example, a solid interface can cause the group of magnetic beads to fall apart, limiting the ability to measure the AMBR signal in a single well. However, by use of an array of magnets to pull the magnetic beads into a critical position, magnetic bead groups formed, stayed together as a unit during rotation, and provided signal on a stand-alone prototype system. In previous work with self-assembled AMBR sensors, hanging drops were used to keep the magnetic bead groups from breaking apart. Being able to use a solid interface (e.g. a standard well plate) have several advantages over previous embodiments: the ability to use large volumes; lack of evaporation; robustness; automation friendliness; the ability to introduce additional media, antimicrobials or other agents; pathogen containment; use of off-the-shelf disposables; and inexpensive design.

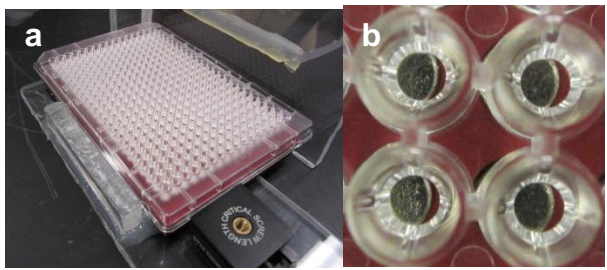

**Fig. S3.** (a) Image of a 384-microwell plate on the magnet array while forming self-assembled AMBR groups in the wells, (b) and a close-up of the wells showing the intentional partial misalignment between the wells and the magnets.

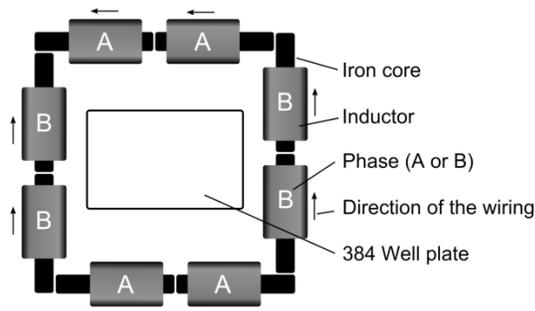

**Fig. S4.** Top view of the inductor setup used to create the rotating magnetic field across 384-microwell plate in the prototype device.

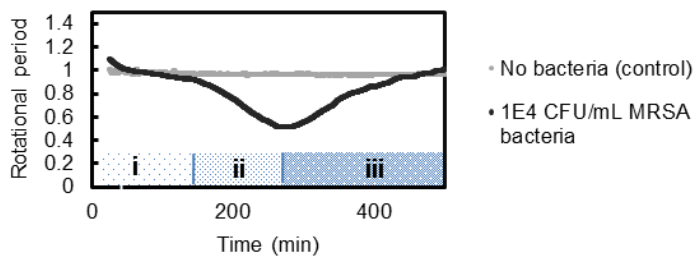

**Fig. S5.** Self-assembled AMBR results with methicillin resistant *S. aureus* (MRSA). The same phases i, ii and iii, seen with *E. coli* in Fig. 3a, are present: initial stable rotation, followed by decrease in rotational period and finally increase in rotational period. The presence of the bacteria can first be seen at roughly 200 minutes from the beginning of the experiment, not including

40 minute preparation time. Control (with no bacteria present) did not show significant changes in the rotation period. The experimental procedure was the same as described for *E. coli* in section 2.